# Bisimulations for Nondeterministic Labeled Markov Processes*

Pedro R. D'Argenio[1,2,3], P. Sánchez Terraf[1,2,3], and Nicolás Wolovick[1]

[1]*FaMAF, Universidad Nacional de Córdoba,* [2]*CONICET and* [3]*CIEM Ciudad Universitaria, 5000 – Córdoba, Argentina.*

**Abstract**

We extend the theory of labeled Markov processes with *internal* nondeterminism, a fundamental concept for the further development of a process theory with abstraction on nondeterministic continuous probabilistic systems. We define *nondeterministic labeled Markov processes (NLMP)* and provide three definition of bisimulations: a bisimulation following a traditional characterization, a *state* based bisimulation tailored to our "measurable" non-determinism, and an *event* based bisimulation. We show the relation between them, including that the largest state bisimulation is also an event bisimulation. We also introduce a variation of the Hennessy-Milner logic that characterizes event bisimulation and that is sound w.r.t. the other bisimulations for arbitrary NLMP. This logic, however, is infinitary as it contains a denumerable ∨. We then introduce a finitary sublogic that characterize all bisimulations for image finite NLMP whose underlying measure space is also analytic. Hence, in this setting, all notions of bisimulation we deal with turn out to be equal. Finally, we show that all notions of bisimulations are different in the general case. The counterexamples that separate them turn to be *non-probabilistic* NLMP.

## 1 Introduction

Markov processes with continuous-state spaces or continuous time evolution (or both) arise naturally in several fields of physics, biology, economics, and computer science (Danos et al. 2006). Many formal frameworks have been defined to study them from a process theory or process algebra perspective (see Strulo 1993; Desharnais 1999; D'Argenio 1999; Bravetti 2002; Desharnais et al. 2002; Bravetti and D'Argenio 2004; D'Argenio and Katoen 2005; Cattani 2005; Cattani et al. 2005; Danos et al. 2006). A prominent and extensive work on this area is the one that builds on top of the so called labeled Markov processes (LMP) (Desharnais 1999; Desharnais et al. 2002). This is due to its solid and well understood mathematical foundations. A LMP allows for many transition probability functions (or Markov kernels) leaving each state (instead of only one as in usual Markov processes). Each transition probability function is a measure ranging on a (possibly continuous) measurable space, and the different transition probability functions can be singled out through labels. Thus this model *does not* consider *internal* nondeterminism. From the process algebra point of view, this is a significant drawback for this theory since internal nondeterminism immediately arises in the analysis of systems, e.g., because of abstracting internal activity (such as weak bisimulation (Milner 1989)) or because of state abstraction techniques (such as in model checking (Clarke et al. 1999)).

*Supported by ANPCyT PICT 26135, SeCyT-UNC, and CONICET

Many other works defined variants of continuous Markov processes that include internal nondeterminism and are mainly used as the underlying semantics of a process algebra (Strulo 1993; D'Argenio 1999; Bravetti 2002; Bravetti and D'Argenio 2004; D'Argenio and Katoen 2005). They also defined a continuous probabilistic variant of the (strong) bisimulation. As correctly pointed out in (Cattani 2005; Cattani et al. 2005), these models lack enough structure to ensure that bisimilar models share the same observable behavior. (This is due to the case in which two objects may be bisimilar but in one of them it is not possible to define probabilistic executions since the transition relation is not a measurable object.) The solution proposed in (Cattani 2005; Cattani et al. 2005) deals with the same unstructured type of models and lift the burden of checking measurability to the semantic tools (such as bisimulation or schedulers). In particular, this results in the definition of a bisimulation as a relation between measures rather than states.

A somewhat related observation has been made by Danos et al. (2006) with respect to the bisimulation relation on LMPs (Desharnais 1999; Desharnais et al. 2002). Danos et al. (2006) show that there are bisimulation relations that may distinguish *beyond events*. That is, states that cannot be *separated* (i.e., distinguished) by any *measurable set* (i.e., any event) may not be related for some bisimulation relation. This is also awkward as events (measurable sets) are the building blocks of observations (probabilistic executions). To overcome this, Danos et al. (2006) define the so called *event bisimulation* (in opposition to the previous *state* bisimulation—name which we will use from now on). An event bisimulation is a sub $\sigma$-algebra $\Lambda$ on the set of states such that the original transition probability functions are also Markov kernels on $\Lambda$, i.e., the original LMP is also an LMP over $\Lambda$. $\Lambda$ induces an equivalence relation $\mathcal{R}(\Lambda)$ also called event bisimulation. Fortunately, it turns out that the largest state bisimulation is also an event bisimulation.

In this paper, we follow the LMP approach towards defining a theory of LMP with internal nondeterminism. Thus, we introduce *nondeterministic labeled Markov processes (NLMP)*. A NLMP has a nondeterministic transition function $T_a$ for each label $a$ that, given a state, it returns a measurable *set* of probability measures (rather than only one probability measure as in LMPs). Moreover, $T_a$ should be measurable. This calls for a definition of a $\sigma$-algebra on top of Giry's $\sigma$-algebra on the set of probability measures (Giry 1981), which we also provide. We give a definition for event bisimulation and state bisimulation and prove similar properties to (Danos et al. 2006), including that the largest state bisimulation is also an event bisimulation. We also provide a definition of "traditional" bisimulation that follows the lines of (Strulo 1993; D'Argenio 1999; Bravetti 2002; D'Argenio and Katoen 2005). We prove that a traditional bisimulation is also a state bisimulation and give sufficient conditions so that the converse holds. Besides, we show that LMPs are just NLMPs without internal nondeterminism and that state (resp. event) bisimulation in the different models agree.

Behavioral equivalences like bisimulation have been characterized using logic with modal-ities, notably the Hennessy-Milner logic (see e.g. van Glabeek 2001). We define an extension of the logic presented in the context of LMP (Desharnais 1999). In fact, the logic is similar to that of Parma and Segala (2007), which was introduced in a discrete setting. However, unlike Parma and Segala (2007), we consider two different formula levels: one that is interpreted on states, and the other that is interpreted on measures. Such separation gives a particular insight: the actual complexity of the model lies exactly on the internal non-determinism introduced by the values of $T_a$ (which are sets of measures). At state level, the logic is as simple as that of Desharnais (1999). We show that this logic completely characterizes event bisimulation and, as a consequence, it is sound w.r.t. traditional and state bisimulation.

In addition, we show that a sublogic of the previous logic characterizes all three bisimulations (event, state and traditional) provided certain restrictions apply, namely, NLMPs are image finite and the state space is analytic. Therefore, all bisimulation equivalences as well as logical equivalence turn out to be the same on this setting.

Nonetheless, we also show that they are different in a more general setting. In the last part of this article, we present two counterexamples, one showing that traditional bisimulation is strictly finer that state and event bisimulation and the other that state bisimulation is strictly finer than event bisimulation. Both counterexamples turn to be *non-probabilistic* NLMPs —which can be seen as a measure theoretic version of labelled transition systems. The first example shows that traditional bisimulation distinguish beyond measurability, and the second one, that event bisimulation has some weakness that has to be overcome.

This article revise and extends our result in (D'Argenio et al. 2009). In particular, Sec. 5, is new to this paper. Most importantly, the new counterexamples presented here lead to new and different conclusions to that of (D'Argenio et al. 2009).

## 2 Fundamentals and Background

In this section we review some foundational theory and prove few basic results that will be of use throughout the paper.

### 2.1 Measure theory

Given a set $S$ and a collection $\Sigma$ of subsets of $S$, we call $\Sigma$ a *$\sigma$-algebra* iff $S \in \Sigma$ and $\Sigma$ is closed under complement and denumerable union. By $\sigma(\mathcal{G})$ we denote the *$\sigma$-algebra generated by* the family $\mathcal{G} \subseteq 2^S$, i.e., the minimal $\sigma$-algebra containing $\mathcal{G}$. Each element of $\mathcal{G}$ is called *generator* and $\mathcal{G}$, the *set of generators*. We call the pair $(S, \Sigma)$ a *measurable space*. A *measurable set* is a set $Q \in \Sigma$. A $\sigma$-additive function $\mu : \Sigma \to [0,1]$ such that $\mu(S) = 1$ is called *probability measure*. By $\delta_a$ we denote the Dirac probability measure concentrated in $\{a\}$. Let $\Delta(S)$ denote the *set of all probability measures* over the measurable space $(S, \Sigma)$. Let $(S_1, \Sigma_2)$ and $(S_1, \Sigma_2)$ be two measurable spaces. A function $f : S_1, \to S_2$ is said to be *measurable* if $\forall Q_2 \in \Sigma_2$, $f^{-1}(Q_2) \in \Sigma_1$, i.e., the inverse function maps measurable sets to measurable sets. In this case we denote $f : (S_1, \Sigma_2) \to (S_1, \Sigma_2)$.

A function $f : S_1 \times \Sigma_2 \to [0,1]$ is a *transition probability* (also called *Markov kernel*) if for all $\omega_1 \in S_1$, $f(\omega_1, \cdot)$ is a probability measure on $(S_2, \Sigma_2)$ and for all $Q_2 \in \Sigma_2$, $f(\cdot, Q_2)$ is measurable.

There is a standard construction by Giry (1981) to endow $\Delta(S)$ with a $\sigma$-algebra as follows: $\Delta(\Sigma)$ is defined as the $\sigma$-algebra generated by the sets of probability measures $\Delta^B(Q) \doteq \{\mu \mid \mu(Q) \in B\}$, with $Q \in \Sigma$ and $B \in \mathcal{B}([0,1])$. ($\mathcal{B}([0,1])$ is the Borel $\sigma$-algebra on the interval $[0,1]$ generated by the open sets.) When $0 \leq p \leq 1$, we will write $\Delta^{\geq p}(Q)$, $\Delta^{>p}(Q)$, $\Delta^{<p}(Q)$, etc. for $\Delta^B(Q)$ with $B = [p,1], (p,1], [0,p)$, etc. respectively. It is known that the set $\{\Delta^{\geq p}(Q) \mid p \in (\mathbb{Q} \cap [0,1]), Q \in \Sigma\}$ generates all $\Delta(\Sigma)$.

On this setting, $f : S_1 \times \Sigma_2 \to [0,1]$ is a transition probability if and only if its curried version $f : S_1 \to \Delta(S_2)$ is measurable. (Mind the notation overloading on $f$.) This follows from the next lemma.

**Lemma 2.1.** $f : S_1 \to \Delta(S_2)$ is measurable iff $f(\cdot, Q) : S_1 \to [0,1]$ is measurable for all $Q \in \Sigma_2$.

*Proof.* It is routine to calculate that $f^{-1}(\Delta^B(Q)) = (f(\cdot, Q))^{-1}(B)$ for all $Q \in \Sigma_2$ and $B \in \mathcal{B}([0,1])$. By this observation, $f^{-1}(\Delta^B(Q)) \in \Sigma_1$ iff $(f(\cdot, Q))^{-1}(B) \in \Sigma_1$. Since it is sufficient to show that $f^{-1}(\Delta^B(Q)) \in \Sigma_1$ for all generators $\Delta^B(Q)$ to state that $f$ is measurable, the lemma follows. □

An important result on Giry's construction is that the $\sigma$-algebra of measures is *separative* (van Breugel 2005), i.e., for any two elements, there is always a measurable set that contains one element but not the other.

**Proposition 2.1.** $\Delta(\Sigma)$ *is separative. That is, given different* $\mu, \mu' \in \Delta(S)$, *there exists* $\Theta \in \Delta(\Sigma)$ *such that* $\mu \in \Theta$ *and* $\mu' \notin \Theta$.

## 2.2 Relations, Measures, and $\sigma$-algebras

Given a relation $R \subseteq S \times S$, the predicate $R\text{-closed}(Q)$ denotes $R(Q) \subseteq Q$. Notice that if $R$ is symmetric, $R\text{-closed}(Q)$ if and only if $\forall s, t : s\,R\,t : s \in Q \Leftrightarrow t \in Q$. Let $(S, \Sigma)$ be a measurable space. For symmetric $R$, define $\Sigma(R) \doteq \{Q \in \Sigma \mid R\text{-closed}(Q)\}$. $\Sigma(R)$ is the sub-$\sigma$-algebra of $\Sigma$ containing all $R$-closed $\Sigma$-measurable sets. The next proposition states that the inclusion order between two relations transfers inversely to the $\sigma$-algebras induced by them and to Giry's construction applied to these $\sigma$-algebras.

**Proposition 2.2.** *Let $R$ and $R'$ be symmetric relations such that $R \subseteq R'$. Then (i) $\Sigma(R) \supseteq \Sigma(R')$ and (ii) $\Delta(\Sigma(R)) \supseteq \Delta(\Sigma(R'))$.*

*Proof.* (i) follows from the fact that any measurable set that is $R'$-closed is also $R$-closed whenever $R \subseteq R'$. For (ii), recall that $\Delta(\Sigma(R'))$ is generated by $\mathcal{G} = \{\Delta^B(Q) \mid Q \in \Sigma(R')$ and $B \in \mathcal{B}([0,1])\}$. Since $\Sigma(R') \subseteq \Sigma(R)$ (by (i)), then $\mathcal{G} \subseteq \Delta(\Sigma(R))$ from which the lemma follows. □

We can lift $R$ to an equivalence relation in $\Delta(S)$ as follows: $\mu R \mu'$ iff $\forall Q \in \Sigma(R) : \mu(Q) = \mu'(Q)$. Then, the predicate $R$-closed can be defined on subsets of $\Delta(S)$ just like before. The following proposition will be useful.

**Proposition 2.3.** *If $R$ is a symmetric relation, every $\Delta(\Sigma(R))$-measurable set is $R$-closed.*

*Proof.* Let $Q \in \Sigma(R)$ and $B \in \mathcal{B}([0,1])$. Then, if $\mu \in \Delta^B(Q)$ and $\mu R \mu'$, $\mu' \in \Delta^B(Q)$. So, each generator $\Delta^B(Q)$ of $\Delta(\Sigma(R))$ is $R$-closed. Moreover, for any symmetric $R$, the property of being $R$-closed is preserved by denumerable union and complement. Since the lifted $R$ is symmetric, we can conclude that every $\Delta(\Sigma(R))$-measurable set is $R$-closed. $\square$

A $\sigma$-algebra $\Sigma$ defines an equivalence relation $\mathcal{R}(\Sigma)$ on $S$ as follows: $s\,\mathcal{R}(\Sigma)\,t$ iff $\forall Q \in \Sigma$, $s \in Q \Leftrightarrow t \in Q$. That is, two elements are related if they cannot be separated by any measurable set. The following properties (due to Danos et al. 2006) appear here for the sake of completeness; they relate $\sigma$-algebras and relations. In particular, (v) is a consequence of (i) and (ii).

**Proposition 2.4.** *Let $(S, \Sigma)$ be a measurable space, $R$ a symmetric relation on $S$, and $\Lambda \subseteq \Sigma$ a sub-$\sigma$-algebra of $\Sigma$. Then, (i) $\Lambda \subseteq \Sigma(\mathcal{R}(\Lambda))$; (ii) $R \subseteq \mathcal{R}(\Sigma(R))$; (iii) if each $R$-equivalence class is in $\Sigma$, then $R = \mathcal{R}(\Sigma(R))$; (iv) $\mathcal{R}(\Lambda) = \mathcal{R}(\Sigma(\mathcal{R}(\Lambda)))$; and (v) $\Sigma(R) = \Sigma(\mathcal{R}(\Sigma(R)))$*[1].

## 2.3 Labeled Markov Processes

A labeled Markov process (LMP) (Desharnais 1999; Desharnais et al. 2002) is a triple $(S, \Sigma, \{\tau_a \mid a \in L\})$ where $\Sigma$ is a $\sigma$-algebra on the set of states $S$, and for each label $a \in L$, $\tau_a : S \times \Sigma \to [0, 1]$ is a transition probability. By Lemma 2.1, we can say that $(S, \Sigma, \{\tau_a \mid a \in L\})$ is an LMP if every $\tau_a : S \to \Delta(S)$ is measurable.

In (Desharnais 1999; Desharnais et al. 2002), a notion of behavioral equivalence similar to Larsen and Skou's (1991) probabilistic bisimulation is introduced.

**Definition 2.1.** *$R \subseteq S \times S$ is a* state bisimulation *on LMP $(S, \Sigma, \{\tau_a \mid a \in L\})$ if it is symmetric*[2] *and for all $s, t \in S$, $a \in L$, $s\,R\,t$ implies that $\tau_a(s)\,R\,\tau_a(t)$.*

This definition is pointwise and not "eventwise" as one should expect in a measure-theoretic realm, besides $R$ has no measurability restriction. In (Danos et al. 2006) a measure-theory aware notion of behavioral equivalence is introduced.

**Definition 2.2.** *An* event bisimulation *on a LMP $(S, \Sigma, \{\tau_a \mid a \in L\})$ is a sub-$\sigma$-algebra $\Lambda$ of $\Sigma$ s.t. $(S, \Lambda, \{\tau_a \mid a \in L\})$ is a LMP.*

Danos et al. (2006) show that $R$ is state bisimulation iff $\Sigma(R)$ is an event bisimulation. This is an important result that leads to prove that the largest state bisimulation is also an event bisimulation (see Theorem 3.4 below).

## 3 Nondeterministic Labeled Markov Processes

In this section we extend the LMP model adding internal nondeterminism. That is, we allow that different but equally labeled transition probabilities leave out the same state. We provide event and state bisimulations for this model, show the relation to LMPs and the relation to earlier definitions of bisimulation on nondeterministic and continuous probabilistic transition systems.

---
[1]Proposition 2.4(v) appears in (Danos et al. 2006) unnecessarily requiring that $R$ is a state bisimulation.
[2]The requirement of symmetry is needed otherwise $\Sigma(R)$ may not be a $\sigma$-algebra.

## 3.1 The model

There have been several attempts to define nondeterministic continuous probabilistic transition systems and all of them are straightforward extensions of (simpler) discrete versions. There are two fundamental differences in our new model. The first one is that the *nondeterministic transition function* $T_a$ now maps states to *measurable sets of* probability measures rather than arbitrary sets as previous approaches. This is motivated by the fact that later on the nondeterminism has to be resolved using schedulers. If we allowed the target set of states to be an arbitrary subset, (as some continuous ones D'Argenio 1999; Bravetti and D'Argenio 2004; Cattani et al. 2005), the system as a whole could suffer from non-measurability issues and therefore it could not be quantified. (Rigorously speaking, labels should also be provided with a $\sigma$-algebra, but we omit it here since it is not needed.) The second difference is inspired by the definition of LMP and Lemma 2.1 (see also the alternative definition of LMP above): we ask that, for each label $a \in L$, $T_a$ is a measurable function. One of the reasons for this restriction is to have well defined modal operators of a probabilistic Hennessy-Milner logic, like in the LMP case.

**Definition 3.1.** *A* nondeterministic labeled Markov process *(NLMP for short) is a structure* $(S, \Sigma, \{T_a \mid a \in L\})$ *where $\Sigma$ is a $\sigma$-algebra on the set of states $S$, and for each label $a \in L$, $T_a : S \to \Delta(\Sigma)$ is measurable.*

For the requirement that $T_a$ is measurable, we need to endow $\Delta(\Sigma)$ with a $\sigma$-algebra. This is a key construction to forthcoming definitions and theorems.

**Definition 3.2.** *$H(\Delta(\Sigma))$ is the minimal $\sigma$-algebra containing all sets $H_\xi \doteq \{\Theta \in \Delta(\Sigma) \mid \Theta \cap \xi \neq \varnothing\}$ with $\xi \in \Delta(\Sigma)$.*

This construction is similar to that of the Effros-Borel spaces (Kechris 1995) and resembles the so-called hit-and-miss topologies (Naimpally 2003). Note that the generator set $H_\xi$ contains all measurable sets that "hit" the measurable set $\xi$. Also observe that $T_a^{-1}(H_\xi)$ is the set of all states $s$ such that, through label $a$, "hit" the set of measures $\xi$ (i.e., $T_a(s) \cap \xi \neq \varnothing$). This forms the basis to existentially quantify over the nondeterminism, and it is fundamental for the behavioral equivalence and the logic.

The next two examples (inspired by an example in Cattani 2005) show why $T_a$ is required to map into measurable sets and to be measurable. For these examples we fix the state space and $\sigma$-algebra in the real unit interval with the standard Borel $\sigma$-algebra.

**Example 3.1.** *Let $\mathcal{V} = \{\delta_q \mid q \in V\}$, where $V$ is the non-measurable Vitali set in $[0, 1]$. It can be shown that $\mathcal{V}$ is not measurable in $\Delta(\Sigma)$. Let $T_a(s) = \mathcal{V}$ for all $s \in [0, 1]$. The resolution of the internal nondeterminism by means of so called schedulers (also adversaries or policies) (Vardi 1985; Puterman 1994), whatever its definition is, would require to assign probabilities to all possible choices. This amounts to measure the nonmeasurable set $T_a(s)$. This is why we require that $T_a$ maps into measurable sets.*

**Example 3.2.** *Let $T_a(s) = \{\mu\}$ for a fixed measure $\mu$, and let $T_b(s) =$ **if** $(s \in V)$ **then** $\{\delta_1\}$ **else** $\varnothing$, for every $s \in [0, 1]$, with $V$ being a Vitali set. Notice that both $T_a(s)$ and $T_b(s)$ are measurable sets for every $s \in [0, 1]$. Supposing that there is a scheduler that chooses to first do $a$ and then $b$ starting at some state $s$, the probability of such set of executions cannot be measured, as it requires to apply $\mu$ to the set $T_b^{-1}(H_{\Delta(S)}) = V$ which is not measurable.*

Besides, we will later need that sets $T_a^{-1}(H_\xi)$ are measurable so that the semantics of the logic maps into measurable sets (see Sec. 4).

## 3.2 NLMPs as a generalization of LMPs

Notice that a LMP is a NLMP without internal nondeterminism. That is, a NLMP in which $T_a(s)$ is a singleton for all $a \in L$ and $s \in S$, is a LMP. In fact, a LMP can be encoded as a NLMP by taking $T_a(s) = \{\tau_a(s)\}$. (We formally prove this in Proposition 3.1.) As a consequence it is necessary that singletons $\{\mu\}$ are measurable in $\Delta(\Sigma)$ for the NLMP to be well defined. The following lemma gives sufficient conditions to ensure that all singletons are measurable in $\Delta(\Sigma)$.

**Lemma 3.1.** *Let $\mathcal{G}$ be a denumerable $\pi$-system on $S$ (i.e., a denumerable subset of $2^S$ containing $S$ and closed under finite intersection). Then, for all $\mu \in \Delta(S)$, $\{\mu\} \in \Delta(\sigma(\mathcal{G}))$.*

*Proof.* It is sufficient to prove that the set

$$\cap \{\Delta^{>q_i}(Q_i) \mid Q_i \in \mathcal{G}, q_i \in \mathbb{Q} \cap [0,1], q_i < \mu(Q_i)\} \cap$$
$$\cap \{\Delta^{<q_i}(Q_i) \mid Q_i \in \mathcal{G}, q_i \in \mathbb{Q} \cap [0,1], \mu(Q_i) < q_i\},$$

which is a denumerable intersection, is equal to the singleton $\{\mu\}$. By construction $\mu$ is in the intersection. Take $\mu'$ s.t. $\mu \neq \mu'$. By a classical theorem of extension of a measure (Billingsley 1995, Theorem 3.3), there must be a $Q_i \in \mathcal{G}$ such that $\mu(Q_i) \neq \mu'(Q_i)$. If $\mu(Q_i) > \mu'(Q_i)$ then $\mu'$ does not belong to the first intersection; if $\mu(Q_i) < \mu'(Q_i)$, $\mu'$ does not belong to the second one. □

In other words, we can guarantee that singletons are measurable in Giry's construction if the underlying $\sigma$-algebra is countably generated.

Note that Lemma 3.1 gives also sufficient conditions to define NLMPs with *finite and denumerable nondeterminism*.

Notice also that asking for measurable singletons in $\Delta(\Sigma)$ does not trivialize $\Sigma$ (in the sense that $\Sigma = 2^S$). A nontrivial example in which Lemma 3.1 holds is the standard Borel $\sigma$-algebra in $\mathbb{R}$. A less obvious example is as follows. Let the $\sigma$-algebra $\mathbf{Q\text{-}coQ} \doteq 2^{\mathbb{Q}} \cup \{\mathbb{R} \setminus \mathbb{Q} \mid Q \in 2^{\mathbb{Q}}\}$. Notice that $\mathbf{Q\text{-}coQ}$ cannot separate one irrational from another (let alone asking for all singletons being measurable). Nevertheless, as it is generated by the denumerable $\pi$-system $\{\{q\} \mid q \in \mathbb{Q}\} \cup \{\varnothing\}$, it is under the conditions of Lemma 3.1 and hence for every measure $\mu$ on it, $\{\mu\}$ is measurable on $\Delta(\mathbf{Q\text{-}coQ})$.

The formal connection between NLMP and LMP is an immediate consequence of the next proposition.

**Proposition 3.1.** *Let $T_a(s) = \{\tau_a(s)\}$ for all $s \in S$ and let $\Sigma$ be a $\sigma$-algebra on $S$. Then $\tau_a : S \to \Delta(S)$ is measurable iff $T_a : S \to \Delta(\Sigma)$ is measurable.*

*Proof.* Let $\xi \in \Delta(\Sigma)$. Note that $T_a(s) \in H_\xi$ iff $\{\tau_a(s)\} \cap \xi \neq \varnothing$ iff $\tau_a(s) \in \xi$. Then $T_a^{-1}(H_\xi) = \tau_a^{-1}(\xi)$. Therefore $\tau_a$ is measurable whenever $T_a$ is measurable. For the converse, we have that $T_a^{-1}(H_\xi)$ is measurable for all generators $H_\xi$. As a consequence $T_a$ is measurable in general. □

## 3.3 The bisimulations

Event bisimulation in NLMP is defined exactly in the same way as for LMP: an event bisimulation is a sub-$\sigma$-algebra that, together with the same set of states and transition of the original NLMP, makes a new NLMP.

**Definition 3.3.** *An* event bisimulation *on a NLMP* $(S, \Sigma, \{T_a \mid a \in L\})$ *is a sub-$\sigma$-algebra* $\Lambda$ *of* $\Sigma$ *s.t.* $T_a : (S, \Lambda) \to (\Delta(\Sigma), H(\Delta(\Lambda)))$ *is measurable for each* $a \in L$.

Notice that $T_a$ is the same function from $S$ to $\Delta(\Sigma)$ only that, for $\Lambda$ to be an event bisimulation, it should be measurable from $\Lambda$ to $H(\Delta(\Lambda))$. Here, $H(\Delta(\Lambda))$ is the sub-$\sigma$-algebra of $H(\Delta(\Sigma))$ generated by $\{H_\xi \mid \xi \in \Delta(\Lambda)\}$.

We extend the notion of event bisimulation to relations. We say that a relation $R$ is an event bisimulation if there is an event bisimulation $\Lambda$ s.t. $R = \mathcal{R}(\Lambda)$. More generally, we say that two states $s, t \in S$ are *event bisimilar*, denoted by $s \sim_e t$, if there is an event bisimulation $\Lambda$ such that $s \, \mathcal{R}(\Lambda) \, t$. The fact that $\sim_e$ is an equivalence relation is an immediate corollary of Theorem 4.5 given below. We remark that, by Proposition 3.1, an event bisimulation on a LMP is also an event bisimulation on the encoding NLMP and vice-versa.

The definition of state bisimulation is less standard. Following the original definition of Milner (1989) (which was lifted to discrete probabilistic models by Larsen and Skou 1991), a traditional definition of bisimulation (see Def. 3.5) verifies that, whenever $s \, R \, t$, every measure on $T_a(s)$ has a corresponding one (modulo $R$) in $T_a(t)$. Rather than looking pointwise at probability measures, our definition follows the idea of Def. 3.2 and verifies that both $T_a(s)$ and $T_a(t)$ *hit* the same measurable sets of measures.

**Definition 3.4.** *A relation* $R \subseteq S \times S$ *is a* state bisimulation *if it is symmetric and for all* $a \in L$, $s \, R \, t$ *implies* $\forall \xi \in \Delta(\Sigma(R)) : T_a(s) \cap \xi \neq \varnothing \Leftrightarrow T_a(t) \cap \xi \neq \varnothing$.

The following property, which also holds in LMPs, states the fundamental relation between state bisimulation and event bisimulation.

**Lemma 3.2.** *Provided $R$ is symmetric, $R$ is a state bisimulation iff $\Sigma(R)$ is an event bisimulation.*

*Proof.* By Def. 3.3, $\Sigma(R)$ is an event bisimulation iff $T_a$ is $\Sigma(R)$-measurable. Since $T_a$ is $\Sigma$-measurable, it suffices to prove that $T_a^{-1}(H_\xi)$ is $R$-closed for all labels $a \in L$ and generators $H_\xi, \xi \in \Delta(\Sigma(R))$.

$R$-closed($T_a^{-1}(H_\xi)$)

iff  ($R$ is symmetric)

$s \, R \, t \Rightarrow \left( s \in T_a^{-1}(H_\xi) \Leftrightarrow t \in T_a^{-1}(H_\xi) \right)$

iff  (Def. inverse function)

$s \, R \, t \Rightarrow (T_a(s) \in H_\xi \Leftrightarrow T_a(t) \in H_\xi)$

iff  (Def. of $H_\xi$)

$s \, R \, t \Rightarrow (T_a(s) \cap \xi \neq \varnothing \Leftrightarrow T_a(t) \cap \xi \neq \varnothing)$.

The last statement is the definition of state bisimulation. □

The following results are consequences of Proposition 2.4 and, for the case of Lemma 3.3.3, Lemma 3.2 and the fact that $\mathcal{R}(\Lambda)$ is an equivalence relation. The proofs are the same as the proofs of similar results for LMP in (Danos et al. 2006).

**Lemma 3.3.** *Let $R$ be a state bisimulation. Then:*
1. *$R$ is an event bisimulation iff $R = \mathcal{R}(\Sigma(R))$.*
2. *If the equivalence classes of $R$ are in $\Sigma$, $R$ is an event bisimulation.*
3. *$\mathcal{R}(\Sigma(R))$ is both a state bisimulation and an event bisimulation.*

Let $\sim_s = \bigcup \{R \mid R$ is a state bisimulation$\}$. In the following we show that $\sim_s$ is also a state bisimulation and hence the largest one. Moreover, we show that $\sim_s$ is also an event bisimulation and, as a consequence, an equivalence relation.

**Theorem 3.4.** *$\sim_s$ is (i) the largest state bisimulation, (ii) an event bisimulation (and hence $\sim_s \subseteq \sim_e$), and (iii) an equivalence relation.*

*Proof.* (i) Take $s, t \in S$ s.t. $s \sim_s t$. Then there is a state bisimulation $R$ with $s R t$. Take a measurable set $\xi \in \Delta(\Sigma(\sim_s))$. Since $R \subseteq \sim_s$, by Proposition 2.2, $\Delta(\Sigma(R)) \supseteq \Delta(\Sigma(\sim_s))$. Hence $\xi \in \Delta(\Sigma(R))$ and by Def. 3.4, $T_a(s) \cap \xi \neq \emptyset \Leftrightarrow T_a(t) \cap \xi \neq \emptyset$ which prove that $\sim_s$ is a state bisimulation. By definition, it is the largest one.
(ii) Because $\sim_s$ is a state bisimulation, $\mathcal{R}(\Sigma(\sim_s))$ is a state bisimulation and an event bisimulation (Lemma 3.3.3). Since $\sim_s$ is the largest bisimulation then $\sim_s = \mathcal{R}(\Sigma(\sim_s))$ and hence it is an event bisimulation.
(iii) By definition, every event bisimulation is an equivalence relation. $\square$

## 3.4 A traditional view to bisimulation

We have already stated that our definition of state bisimulation differs from a more traditional view such as those in (Strulo 1993; D'Argenio 1999; Bravetti 2002; D'Argenio and Katoen 2005; Bravetti and D'Argenio 2004). These definitions closely resemble Larsen and Skou's (1991) definition. (The only difference is that two measures are considered equivalent if they agree in every *measurable union of* equivalence classes induced by the relation.) In the following, we give a more "modern" variant of this definition.

**Definition 3.5.** *A relation $R$ is a* traditional bisimulation *if it is symmetric and for all $a \in L$, $s R t$ implies $T_a(s) R T_a(t)$. We say that $s, t \in S$ are* traditionally bisimilar, *denoted by $s \sim_t t$, if there is an traditional bisimulation $R$ such that $s R t$.*

Note that $R$ is lifted this time to sets as is usual: $T_a(s) R T_a(t)$ if for all $\mu \in T_a(s)$, there is $\mu' \in T_a(t)$ s.t. $\mu R \mu'$ and vice-versa. (Had we explicitly written this definition on Def. 3.5, it would have resembled traditional definitions.)

The proof of the next proposition follows the standard strategy of the classic bisimulation (see Milner 1989). Apart from the probabilistic treatment, it only differs in that the composition $R \circ R'$ is granted to be traditional bisimulation if $R$ and $R'$ are *reflexive* traditional bisimulations. (If one of $R$ or $R'$ is not reflexive, $R \circ R'$ may not be a traditional bisimulation.)

**Proposition 3.2.** *$\sim_t$ is a traditional bisimulation and an equivalence relation.*

In the following we discuss the relation between state bisimulation and traditional bisimulation. Lemma 3.5 states that every traditional bisimulation is a state bisimulation. Theorems 3.6 and 3.7 give sufficient conditions to strengthen Lemma 3.5 so that the converse also holds.

**Lemma 3.5.** *If $R$ is a traditional bisimulation, then $R$ is a state bisimulation.*

*Proof.* Let $s\ R\ t$ and $\xi \in \Delta(\Sigma(R))$. If $T_a(s) \cap \xi \neq \varnothing$, then there is $\mu \in T_a(s)$ s.t. $\mu \in \xi$. Since $R$ is a traditional bisimulation, $T_a(s)\ R\ T_a(t)$, i.e., there is $\mu' \in T_a(t)$ s.t. $\mu R \mu'$. By Proposition 2.3 $R$-closed($\xi$), so $\mu' \in \xi$, and hence $T_a(t) \cap \xi \neq \varnothing$ as required. The other implication follows by symmetry. □

In the following we give two sufficient conditions so that a state bisimulation is also a traditional bisimulation. The first condition focuses on the NLMP. It requires the NLMP to be *image denumerable*.

**Definition 3.6.** *A NLMP $(S, \Sigma, \{T_a \mid a \in L\})$ is* image denumerable *iff for all $a \in L, s \in S$, $T_a(s)$ is denumerable.*

**Theorem 3.6.** *Let $(S, \Sigma, \{T_a \mid a \in L\})$ be an image denumerable NLMP. Then $R$ is a traditional bisimulation iff it is a state bisimulation.*

*Proof.* The left-to-right implication is Lemma 3.5. For the other implication we proceed as follows.

Let $s\ R\ t$ and for all $\xi \in \Delta(\Sigma(R))$, $T_a(s) \cap \xi \neq \varnothing \Leftrightarrow T_a(t) \cap \xi \neq \varnothing$. Suppose towards a contradiction that $T_a(s)\ \not R\ T_a(t)$, i.e. $\exists \mu \in T_a(s), \forall \mu'_i \in T_a(t) : \exists Q_i \in \Sigma(R) : \mu(Q_i) \bowtie_i \mu'_i(Q_i)$, where $\bowtie_i \in \{>, <\}$ and $i \in \mathbb{N}$ (the NLMP is image denumerable). By density of the rationals, there are $\{q_i\}_i \subseteq \mathbb{Q} \cap [0,1]$ such that $\mu(Q_i) \bowtie_i q_i \bowtie_i \mu'_i(Q_i)$. Then $\mu \in \Delta^{\bowtie_i q_i}(Q_i) \not\ni \mu'_i$. Let $\xi \doteq \cap_i \Delta^{\bowtie_i q_i}(Q_i)$. This set is measurable, moreover, since every $Q_i \in \Sigma(R)$, so $\xi \in \Delta(\Sigma(R))$. Then $\mu \in T_a(s) \cap \xi$, but $T_a(t) \cap \xi = \varnothing$ hence contradicting the assumption. □

After reading the proof, it should be clear that we can relax the sufficient condition to require that the partition $T_a(s)/R$ is denumerable for each state $s$ and label $a$ instead of image denumerability.

Observe that a state bisimulation on a LMP is a traditional bisimulation on the encoding NLMP and vice-versa since $\{\tau_a(s)\} = T_a(s)\ R\ T_a(t) = \{\tau_a(t)\}$ iff $\tau_a(s)\ R\ \tau_a(t)$. As a consequence of Lemma 3.5 and Theorem 3.6 (a deterministic NLMP is image denumerable!), we conclude that a state bisimulation on a LMP is a state bisimulation on the encoding NLMP and vice-versa.

The second sufficient condition looks at the $\sigma$-algebra $\Sigma(R)$ induced by the state bisimulation $R$. It turns out that if $\Sigma(R)$ is generated by a denumerable $\pi$-system, $R$ is also a traditional bisimulation.

**Theorem 3.7.** *Let $R$ be a symmetric relation such that $\Sigma(R)$ is generated by a denumerable set $\mathcal{G}$. Then $R$ is a traditional bisimulation iff it is a state bisimulation.*

*Proof.* As before, the left-to-right implication is Lemma 3.5. For the other implication we proceed as follows. Suppose towards a contradiction that $s \, R \, t$ and $T_a(s) \, \cancel{R} \, T_a(t)$, i.e. $\exists \mu \in T_a(s), \forall \mu' \in T_a(t) : \mu \, \cancel{R} \, \mu'$. By (Billingsley 1995, Theorem 3.3), this implies that there exists $Q_i \in \pi(\mathcal{G})$ s.t. $\mu(Q_i) \neq \mu'(Q_i)$ with $i \in \mathbb{N}$. (Notice that $\pi(\mathcal{G})$, the $\pi$-system generated by $\mathcal{G}$, is also denumerable and generates $\Sigma(R)$.) The rest of the proof is as in Theorem 3.6. □

## 4 A Logic for Bisimulation on NLMP

The logic we present below is based on the logic given by Parma and Segala (2007). The main difference is that we consider two kinds of formulas: one that is interpreted on states, and another that is interpreted on measures. The syntax is as follows,

$$\varphi \equiv \top \mid \varphi_1 \wedge \varphi_2 \mid \langle a \rangle \psi$$
$$\psi \equiv \bigvee_{i \in I} \psi_i \mid \neg \psi \mid [\varphi]_{\geq q}$$

where $a \in L$, $I$ is a denumerable index set, and $q \in \mathbb{Q} \cap [0,1]$. We denote by $\mathcal{L}$ the set of all formulas generated by the first production and by $\mathcal{L}_\Delta$ the set of all formulas generated by the second production.

The semantics is defined with respect to a NLMP $(S, \Sigma, T)$. Formulas in $\mathcal{L}$ are interpreted as sets of states in which they become true, and formulas in $\mathcal{L}_\Delta$ are interpreted as sets of measures on the state space as follows,

$$[\![\top]\!] = S \qquad\qquad [\![\bigvee_{i \in I} \psi_i]\!] = \bigcup_i [\![\psi_i]\!]$$
$$[\![\varphi_1 \wedge \varphi_2]\!] = [\![\varphi_1]\!] \cap [\![\varphi_2]\!] \qquad\qquad [\![\neg \psi]\!] = [\![\psi]\!]^c$$
$$[\![\langle a \rangle \psi]\!] = T_a^{-1}(H_{[\![\psi]\!]}) \qquad\qquad [\![[\varphi]_{\geq q}]\!] = \Delta^{\geq q}([\![\varphi]\!])$$

In particular, notice that $\langle a \rangle \psi$ is valid in a state $s$ whenever there is some measure $\mu \in T_a(s)$ that makes $\psi$ valid, and that $[\varphi]_{\geq q}$ is valid in a measure $\mu$ whenever $\mu([\![\varphi]\!]) \geq q$. As a consequence, we need that sets $[\![\varphi]\!]$ and $[\![\psi]\!]$ are measurable in $\Sigma$ and $\Delta(\Sigma)$, respectively. Indeed, this follows straightforwardly by induction on the construction of the formula after observing that all operations involved in the definition of the semantics preserve measurability (in particular $T_a$ is a measurable function). For the rest of the section, fix $[\![\mathcal{L}]\!] = \{[\![\varphi]\!] \mid \varphi \in \mathcal{L}\}$ and $[\![\mathcal{L}_\Delta]\!] = \{[\![\psi]\!] \mid \psi \in \mathcal{L}_\Delta\}$.

We particularly notice that some other operators can be encoded as syntactic sugar. For instance, we can define $[\varphi]_{>r} \equiv \bigvee_{q \in \mathbb{Q} \cap [0,1] \wedge q > r} [\varphi]_{\geq q}$ for any real $r \in [0,1]$, and $[\varphi]_{\leq r} \equiv \neg [\varphi]_{>r}$.

We show that $\mathcal{L}$ characterizes event bisimulation. This is an immediate consequence of the fact that $\sigma([\![\mathcal{L}]\!])$, the $\sigma$-algebra generated by the logic $\mathcal{L}$, is the smallest event bisimulation, which is what we aim to prove in this part of the section. The proof strategy resembles that of (Danos et al. 2006, Sec. 5) but it is properly tailored to our two level logic. Moreover, such a separation allowed us to find an alternative to Dynkin's Theorem (used in Danos et al. 2006).

We extend the definition of $\Delta(\mathcal{C})$ to any arbitrary set $\mathcal{C} \subseteq \Sigma$ by taking $\Delta(\mathcal{C})$ to be the $\sigma$-algebra generated by $\Delta^{\geq p}(Q)$ with $Q \in \mathcal{C}$ and $p \in [0,1]$. From now on we write $\sigma(\mathcal{L})$, $\Delta(\mathcal{L})$ and $\mathcal{R}(\mathcal{L})$ instead of $\sigma([\![\mathcal{L}]\!])$, $\Delta([\![\mathcal{L}]\!])$ and $\mathcal{R}([\![\mathcal{L}]\!])$, respectively.

The concept of *stable family* of measurable sets is crucial to the proof of Theorem 4.5.

**Definition 4.1.** *Given a NLMP $(S, \Sigma, T)$, the family $\mathcal{C} \subseteq \Sigma$ is* stable *for $(S, \Sigma, T)$ if for all $a \in L$ and $\xi \in \Delta(\mathcal{C})$, $T_a^{-1}(H_\xi) \in \mathcal{C}$.*

Notice that $\mathcal{C}$ is an event bisimulation iff it is a stable $\sigma$-algebra.

The key point of the proof is to show that $[\![\mathcal{L}]\!]$ is the smallest stable $\pi$-system, which is stated in Lemma 4.2. The next lemma is auxiliary to Lemma 4.2.

**Lemma 4.1.** $[\![\mathcal{L}_\Delta]\!] = \Delta(\mathcal{L})$

*Proof.* $[\![\mathcal{L}_\Delta]\!]$ is a $\sigma$-algebra since: *(i)* $\Delta(S) = [\![[\top]_{\geq 1}]\!] \in [\![\mathcal{L}_\Delta]\!]$; *(ii)* for $\xi_i \in [\![\mathcal{L}_\Delta]\!]$ there are $\psi_i \in \mathcal{L}_\Delta$ s.t. $\xi_i = [\![\psi_i]\!]$, and hence $\bigcup_i \xi_i = \bigcup_i [\![\psi_i]\!] = [\![\bigvee_{i \in I} \psi_i]\!] \in [\![\mathcal{L}_\Delta]\!]$; and *(iii)* for $\xi \in [\![\mathcal{L}_\Delta]\!]$ there is $\psi \in \mathcal{L}_\Delta$ s.t. $\xi = [\![\psi]\!]$, and hence $\xi^{\mathsf{c}} = [\![\psi]\!]^{\mathsf{c}} = [\![\neg \psi]\!] \in [\![\mathcal{L}_\Delta]\!]$. Moreover, since $[\![[\varphi]_{\geq p}]\!] = \Delta^{\geq p}([\![\varphi]\!])$, every generator set of $\Delta(\mathcal{L})$ is in $[\![\mathcal{L}_\Delta]\!]$ and hence $\Delta(\mathcal{L}) \subseteq [\![\mathcal{L}_\Delta]\!]$.

Finally, it can be proven by induction on the depth of the formula that $[\![\mathcal{L}_\Delta]\!] \subseteq \mathcal{C}$ for any $\sigma$-algebra $\mathcal{C}$ containing all sets $[\![[\varphi]_{\geq p}]\!] = \Delta^{\geq p}([\![\varphi]\!])$ for $p \in [0,1]$ and $\varphi \in \mathcal{L}$. Then $[\![\mathcal{L}_\Delta]\!]$ is the smallest $\sigma$-algebra containing all generator sets of $\Delta(\mathcal{L})$. Therefore $[\![\mathcal{L}_\Delta]\!] = \Delta(\mathcal{L})$. □

**Lemma 4.2.** $[\![\mathcal{L}]\!]$ *is the smallest stable $\pi$-system for $(S, \Sigma, T)$.*

*Proof.* $[\![\mathcal{L}]\!]$ is a $\pi$-system since: *(i)* $S = [\![\top]\!] \in [\![\mathcal{L}]\!]$ and *(ii)* for $Q_1, Q_2 \in [\![\mathcal{L}]\!]$ there are $\varphi_1, \varphi_2 \in \mathcal{L}$ s.t. $Q_1 = [\![\varphi_1]\!]$ and $Q_2 = [\![\varphi_2]\!]$, and hence $Q_1 \cap Q_2 = [\![\varphi_1]\!] \cap [\![\varphi_2]\!] = [\![\varphi_1 \wedge \varphi_2]\!] \in [\![\mathcal{L}]\!]$.

For stability, let $\xi \in \Delta(\mathcal{L})$. By Lemma 4.1, there is $\psi \in \mathcal{L}_\Delta$ s.t. $[\![\psi]\!] = \xi$. Then $T_a^{-1}(H_\xi) = T_a^{-1}(H_{[\![\psi]\!]}) = [\![\langle a \rangle \psi]\!] \in [\![\mathcal{L}]\!]$.

Let $\mathcal{C}$ be another stable $\pi$-system for $(S, \Sigma, T)$. By induction in the depth of the formula we show simultaneously that $\mathcal{C} \supseteq [\![\mathcal{L}]\!]$ and $\Delta(\mathcal{C}) \supseteq \Delta(\mathcal{L})$. First notice that $[\![\top]\!] = S \in \mathcal{C}$ since $\mathcal{C}$ is a $\pi$-system. Now, suppose inductively that $[\![\varphi]\!], [\![\varphi_1]\!], [\![\varphi_2]\!] \in \mathcal{C}$ and $[\![\psi]\!], [\![\psi_i]\!] \in \Delta(\mathcal{C})$ for $i \geq 0$. Then: *(i)* $[\![\varphi_1 \wedge \varphi_2]\!] = [\![\varphi_1]\!] \cap [\![\varphi_2]\!] \in \mathcal{C}$, because $\mathcal{C}$ is a $\pi$-system; *(ii)* $[\![\langle a \rangle \psi]\!] = T_a^{-1}(H_{[\![\psi]\!]}) \in \mathcal{C}$, because $\mathcal{C}$ is stable; *(iii)* $[\![\bigvee_{i \in I} \psi_i]\!] = \bigcup_i [\![\psi_i]\!] \in \Delta(\mathcal{C})$ and *(iv)* $[\![\neg \psi]\!] = [\![\psi]\!]^{\mathsf{c}} \in \Delta(\mathcal{C})$ because $\Delta(\mathcal{C})$ is a $\sigma$-algebra; and finally, *(v)* $[\![[\varphi]_{\geq p}]\!] = \Delta^{\geq p}([\![\varphi]\!]) \in \Delta(\mathcal{C})$ by definition of generator set of $\Delta(\mathcal{C})$. □

Lemma 4.3 is auxiliary to Lemma 4.4. It is also significantly simpler than its relative in (Danos et al. 2006, Lemma 5.4). This is due to our definition of stability and the use of a powerful result of Viglizzo (2005).

**Lemma 4.3.** *If $\mathcal{C}$ is a stable $\pi$-system for $(S, \Sigma, T)$, then $\sigma(\mathcal{C})$ is also stable.*

*Proof.* First notice that $\mathcal{C}$ is stable iff $\{T_a^{-1}(H_\xi) \mid a \in L, \xi \in \Delta(\mathcal{C})\} \subseteq \mathcal{C}$. By (Viglizzo 2005, Lemma 3.6), $\Delta(\mathcal{C}) = \Delta(\sigma(\mathcal{C}))$. Then $\{T_a^{-1}(H_\xi) \mid a \in L, \xi \in \Delta(\sigma(\mathcal{C}))\} \subseteq \mathcal{C} \subseteq \sigma(\mathcal{C})$, which proves that $\sigma(\mathcal{C})$ is stable. □

The next lemma is central to the proof that $\mathcal{L}$ characterizes event bisimulation, which is then presented in Theorem 4.5.

**Lemma 4.4.** *$\sigma(\mathcal{L})$ is the smallest stable $\sigma$-algebra included in $\Sigma$.*

*Proof.* Let $\mathcal{F}$ be the smallest stable $\sigma$-algebra included in $\Sigma$. By Lemma 4.2, $[\![\mathcal{L}]\!] \subseteq \mathcal{F}$, since $\mathcal{F}$ is a stable $\pi$-system. Therefore $\sigma(\mathcal{L}) \subseteq \mathcal{F}$ since $\mathcal{F}$ is also a $\sigma$-algebra. For the other inclusion, notice that $[\![\mathcal{L}]\!]$ is a stable $\pi$-system because of Lemma 4.2. By Lemma 4.3, $\sigma(\mathcal{L})$ is stable, therefore it contains $\mathcal{F}$. □

**Theorem 4.5.** *The logic $\mathcal{L}$ completely characterizes event bisimulation. In other words, $\mathcal{R}(\mathcal{L}) = \sim_e$*

*Proof.* Lemma 4.4 establishes that $\sigma(\mathcal{L})$ is stable, i.e. it is an event bisimulation. Being the smallest, it implies that any other event bisimulation preserves $\mathcal{L}$ formulas. □

A consequence of this theorem together with Theorem 3.4 and Lemma 3.5 is that both traditional and state bisimulation are sound for $\mathcal{L}$, i.e., they preserve the validity of formulas.

**Theorem 4.6.** $\sim_t \subseteq \sim_s \subseteq \sim_e = \mathcal{R}(\mathcal{L})$.

## 4.1 Completeness on image finite NLMPs

The rest of the section is devoted to show that the logic completely characterizes (all three) bisimulation on NLMPs with image finite nondeterminism and standing on analytic spaces. In fact, we show completeness of the sublogic of $\mathcal{L}$ defined by:

$$\varphi \;\equiv\; \top \;|\; \varphi_1 \wedge \varphi_2 \;|\; \langle a \rangle [\bowtie_i q_i \varphi_i]_{i=1}^n$$

where $\bowtie_i \in \{>, <\}$ and $q_i \in \mathbb{Q} \cap [0,1]$. We define the new modal operation as a shorthand notation: $\langle a \rangle [\bowtie_i q_i \varphi_i]_{i=1}^n \equiv \langle a \rangle \bigwedge_{i=1}^n [\varphi]_{\bowtie_i q_i}$. Therefore, its semantics is given by $[\![\langle a \rangle [\bowtie_i q_i \varphi_i]_{i=1}^n]\!] = T_a^{-1}(H_{\bigcap_{i=1}^n \Delta^{\bowtie_i q_i}([\![\varphi_i]\!])})$. Let $\mathcal{L}_f \subseteq \mathcal{L}$ denote the set of all formulas defined with the grammar above. Notice that $\mathcal{L}_f$ is a denumerable set whenever the set of labels $L$ is denumerable.

The expression $\langle a \rangle [\bowtie_i q_i \varphi_i]_{i=1}^n$ is like a conjunction of formulas $\langle a \rangle_{\bowtie_i q_i} \varphi_i$, but the probabilistic bounds must be satisfied by the *same* nondeterministic transition. Modality $\langle a \rangle_{\bowtie q} \varphi$ suffices to characterize bisimulation on LMP (Desharnais et al. 2002) but, as we see in the next example that originates in (Celayes 2006), it is not enough for the more general setting of NLMPs.

**Example 4.1.** *Take the discrete NLMPs depicted in Fig. 1. States $s$ and $t$ are not bisimilar since given a $\mu \in T_a(s)$, there is no $\mu' \in T_a(t)$ such that $\mu(Q) = \mu'(Q)$ for all $Q \in \{\{x\}, \{y\}, \{z\}\}$ (which are the only relevant possible R-closed sets). A logic having a modality that can only describe one behavior after a label will not be able to distinguish between $s$ and $t$. For example, $[\![\langle a \rangle_{>q} \varphi]\!] = \{w \mid T_a(w) \cap \Delta^{>q}([\![\varphi]\!]) \neq \varnothing\}$ will always have $s$ and $t$ together. Observe that negation, denumerable conjunction or disjunction, do not add any distinguishing power (on an image finite setting).*

The essential need for this new modal operator also shows that our $\sigma$-algebra $H(\Delta(\Sigma))$ in Def. 3.2 can not be simplified to $\sigma(\{H_{\Delta^B(Q)} : B \in \mathcal{B}([0,1]), Q \in \Sigma\})$. States $s$ and $t$ in the example above should be observationally distinguished from each other. Formally, this amounts to say that there must be some label $a$ and some measurable $\Theta$ such that $T_a^{-1}(\Theta)$ separates $\{s\}$ from $\{t\}$. Therefore, the same must be true for some generator $\Theta$, but this does not hold for the family $\{H_{\Delta^B(Q)} : B \in \mathcal{B}([0,1]), Q \in \Sigma\}$.

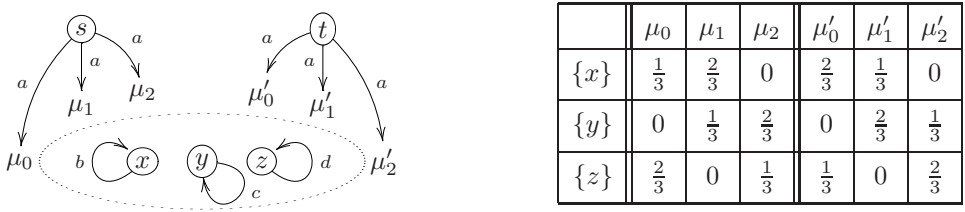

Figure 1: $s$ and $t$ are not bisimilar

Logical characterization of bisimulation is succinctly stated as $s \sim_s t \Leftrightarrow s\ \mathcal{R}(\mathcal{L}_f)\ t$ (similarly for $\sim_t$). The left-to-right implication is immediate by Theorem 4.6. For the converse, we restrict the state space and the branching.

The strategy is to prove that $\mathcal{R}(\mathcal{L}_f)$ is a traditional bisimulation, that is, $s\ \mathcal{R}(\mathcal{L}_f)\ t$ implies that $\forall \mu \in T_a(s), \exists \mu' \in T_a(t), \mu\ \mathcal{R}(\mathcal{L}_f)\ \mu'$; recall this means $\mu(Q) = \mu'(Q)$ for all $Q \in \Sigma(\mathcal{R}(\mathcal{L}_f))$. For analytic spaces this holds if it is valid for the restricted set of $Q \in \Sigma(\mathcal{R}(\mathcal{L}_f))$ such that $Q = [\![\varphi]\!]$, for some $\varphi \in \mathcal{L}_f$. We first introduce analytic spaces and a result from descriptive set theory that is fundamental for the proof.

**Definition 4.2.** *A topological space is* Polish *if it is separable (i.e. it contains a countable dense subset) and completely metrizable. A topological space is* analytic *if it is the continuous image of a Polish space. A measurable space is* analytic (standard) Borel *if it is isomorphic to $(X, \sigma(\mathcal{T}))$ where $\mathcal{T}$ is an analytic (Polish) topology on $X$.*

Every standard Borel space is analytic, but the converse is false. The real line with the usual Borel $\sigma$-algebra, and more generally, $A^{\mathbb{N}}$ with $A$ a countable discrete space, are standard Borel and therefore, analytic.

The next theorem from (Desharnais and Panangaden 2003) essentially shows that in analytic Borel spaces, the $R$-closed measurable sets are well-behaved when the relation $R$ is defined in terms of a sequence of measurable sets.

**Theorem 4.7.** *Let $(S, \Sigma)$ be an analytic Borel space. Let $\mathcal{F} \subseteq \Sigma$ be countable and assume $S \in \mathcal{F}$. Then $\Sigma(\mathcal{R}(\mathcal{F})) = \sigma(\mathcal{F})$.*

The following lemma provides a general framework to prove that a logic characterizes bisimulation. In fact it has been used to prove that less expressive logics characterize traditional bisimulation in some restricted NLMPs (Celayes 2006).

**Lemma 4.8.** *Let $(S, \Sigma, T)$ be a NLMP with $< S, \Sigma >$ being an analytic Borel space. Let $\mathfrak{L}$ be a logic s.t. (i) $\mathfrak{L}$ contains operators $\top$ and $\wedge$ with the usual semantics; (ii) for every formula $\varphi \in \mathfrak{L}$, $[\![\varphi]\!]$ is $\Sigma$-measurable; (iii) the set of all formulas in $\mathfrak{L}$ is denumerable; and (iv) for every $s\ \mathcal{R}(\mathfrak{L})\ t$ and every $\mu \in T_a(s)$ there exists $\mu' \in T_a(t)$ such that $\forall \varphi \in \mathfrak{L}, \mu([\![\varphi]\!]) = \mu'([\![\varphi]\!])$. Then, two logically equivalent states $s, t$ are traditionally bisimilar.*

*Proof.* Let $\mathcal{F} = \{[\![\varphi]\!] \mid \varphi \in \mathfrak{L}\}$. Because of *(i)*, $[\![\top]\!] = S$ and $[\![\varphi_1]\!] \cap [\![\varphi_2]\!] = [\![\varphi_1 \wedge \varphi_2]\!]$. Hence $\mathcal{F}$ forms a $\pi$-system. Because of *(iv)*, $\mu, \mu'$ agree in $\mathcal{F}$ and, by (Billingsley 1995, Theorem 3.3), they also agree in $\sigma(\mathcal{F})$. Notice that hypotheses of Theorem 4.7 are met, i.e., $\Sigma$ is analytic, $\mathcal{F} \subseteq \Sigma$ is countable (by *(ii)* and *(iii)*) such that $S \in \mathcal{F}$ (by *(i)*), and $\mathcal{R}(\mathfrak{L})$ equals $\mathcal{R}(\mathcal{F})$. Therefore, by Theorem 4.7, $\sigma(\mathcal{F}) = \Sigma(\mathcal{R}(\mathfrak{L}))$, which implies that $\mu$ and $\mu'$ agree in $\Sigma(\mathcal{R}(\mathfrak{L}))$. Since $\mathcal{R}(\mathfrak{L})$ is symmetric, $\mathcal{R}(\mathfrak{L})$ is a traditional bisimulation. □

Notice that Lemma 4.8 holds for any logic fulfilling the hypothesis, in particular it should encode the transfer property of the bisimulation and may not contain negation. We already know that $\mathcal{L}_f$ has operators $\top$ and $\wedge$, is denumerable, and that each formula is interpreted in a $\Sigma$-measurable set. In the following, we show that the transfer property can be encoded by using the modality.

**Lemma 4.9.** *Let $(S, \Sigma, T)$ be an image finite NLMP (i.e. $T_a(s)$ is finite for all $a \in L, s \in S$). Then for every pair of states such that $s \, \mathcal{R}(\mathcal{L}_f) \, t$ and $\mu \in T_a(s)$, there is a $\mu' \in T_a(t)$ such that $\forall \varphi \in \mathcal{L}_f, \mu(\llbracket \varphi \rrbracket) = \mu'(\llbracket \varphi \rrbracket)$.*

*Proof.* Suppose towards a contradiction that there are $s, t$ with $s \, \mathcal{R}(\mathcal{L}_f) \, t$ and there is a $\mu \in T_a(s)$, such that for all $\mu'_i \in T_a(t)$ there is a formula $\varphi_i \in \mathcal{L}_f$ with $\mu(\llbracket \varphi_i \rrbracket) \neq \mu'_i(\llbracket \varphi_i \rrbracket)$. Since $T_a(t)$ is finite, there are at most $n$ different $\mu'_i$. We can choose $\bowtie_i \in \{>, <\}, q_i \in \mathbb{Q} \cap [0, 1]$ accordingly to make $\mu(\llbracket \varphi_i \rrbracket) \bowtie_i q_i \bowtie_i \mu'_i(\llbracket \varphi_i \rrbracket)$. Take $\psi = \langle a \rangle [\bowtie_i q_i \varphi_i]_{i=1}^n$. Then $s \in \llbracket \psi \rrbracket$ but $t \notin \llbracket \psi \rrbracket$ contradicting $s \, \mathcal{R}(\mathcal{L}_f) \, t$. □

So, finally, we can state the following theorem.

**Theorem 4.10.** *Let $(S, \Sigma, T)$ be an image finite NLMP with $(S, \Sigma)$ being analytic. For all $s, t \in S$,*

$$s \sim_t t \quad \Leftrightarrow \quad s \sim_s t \quad \Leftrightarrow \quad s \sim_e t \quad \Leftrightarrow \quad s \, \mathcal{R}(\mathcal{L}_f) \, t$$

*Proof.* $s \sim_t t \Rightarrow s \sim_s t \Rightarrow s \sim_e t \Leftrightarrow s \, \mathcal{R}(\mathcal{L}) \, t$ (by Theorem 4.6) $\Rightarrow s \, \mathcal{R}(\mathcal{L}_f) \, t$ (because $\mathcal{L}_f \subseteq \mathcal{L}$) $\Rightarrow s \sim_t t$ (by Lemmas 4.8 and 4.9). □

## 5 Non-probabilistic NLMPs and Counterexamples

The purpose of this section is to construct counterexamples over standard Borel spaces witnessing that all our notions of bisimilarity are different in the case of uncountable nondeterminism. Moreover, it suffices to consider a non-probabilistic variant of NLMP, in which transitions only map into a set of Dirac measures. These structures looks very much like LTSs, the only exception being that the state space has a $\sigma$-algebra attached.

Somehow, the type of counterexamples —non-probabilistic NLMPs over standard Borel spaces with uncountable branching— shows that our Theorems 3.6 and 3.7 are the best possible, even if we assume that our state space is the Borel space of the real numbers.

### 5.1 The subspace of Dirac measures

Since the counterexample NLMPs only run on Dirac measures over standard Borel spaces, we focus first on understanding these objects.

Let $(S, \Sigma)$ be a measurable space. We call $\delta(P) = \{\delta_s : s \in P\}$ for each $P \subseteq S$. The set $\delta(S)$ inherits the measurable structure from $\Delta(S)$ by restriction: its $\sigma$-algebra is

$$\Delta(\Sigma)|\delta(S) = \{\xi \cap \delta(S) : \xi \in \Delta(\Sigma)\}.$$

Notice that the elements of $\Delta(\Sigma)|\delta(S)$ are not necessarily measurable sets in $(\Delta(S), \Delta(\Sigma))$. However it is indeed the case if $\Sigma$ is Borel standard. This is stated in the following proposition.

**Proposition 5.1.** *1. The map $s \mapsto \delta_s$ is a measurable embedding between $S$ and $\Delta(S)$, i.e., the function $\delta_{(\cdot)}$ is a bijection between $S$ and its image $\delta(S)$ s.t. both $\delta_{(\cdot)}$ and $\delta_{(\cdot)}^{-1}$ are measurable.*

*2. If $(S, \Sigma)$ is a standard Borel space, $\delta(S)$ belongs to $\Delta(\Sigma)$, i.e., it is a measurable set and hence $\Delta(\Sigma)|\delta(S) \subseteq \Delta(\Sigma)$.*

*3. If $(S, \Sigma)$ is standard and $X \subseteq S$, $\delta(X)$ is measurable if and only if $X$ is measurable.*

*Proof.* It is clear that $\delta$ is injective. To show it is an embedding amounts to prove that $\Delta(\Sigma)|\delta(S) = \{\delta(Q) : Q \in \Sigma\}$. First observe that $\Delta(\Sigma)$ is the smallest family that contains $\mathcal{G} = \{\Delta^{<q}(Q), (\Delta(S) \setminus \Delta^{<q}(Q)) : q \in \mathbb{Q}, Q \in \Sigma\}$ and is closed under countable intersections and unions. We first show that for every $\xi \in \mathcal{G}$, $\xi \cap \delta(S)$ is of the form $\delta(Q)$:

$$\Delta^{<q}(Q) \cap \delta(S) = \begin{cases} \varnothing & q \leq 0 \\ \delta(S \setminus Q) & 0 < q \leq 1 \\ \delta(S) & q > 1. \end{cases}$$

$$(\Delta(S) \setminus \Delta^{<q}(Q)) \cap \delta(S) = \begin{cases} \delta(S) & q \leq 0 \\ \delta(Q) & 0 < q \leq 1 \\ \varnothing & q > 1. \end{cases}$$

Incidentally, this also proves $\Delta(\Sigma)|\delta(S) \supseteq \{\delta(Q) : Q \in \Sigma\}$. Assume $\xi_i \cap \delta(S)$ is of the form $\delta(Q_i)$ with $Q_i \in \Sigma$ for every $i$. Then

$$\left(\bigcup_i \xi_i\right) \cap \delta(S) = \bigcup_i \xi_i \cap \delta(S) = \bigcup_i \delta(Q_i) = \delta\left(\bigcup_i Q_i\right)$$

Since $\Sigma$ is a $\sigma$-algebra, $\left(\bigcup_i \xi_i\right) \cap \delta(S)$ is of the form $\delta(Q)$. The same works for countable intersections, and we have the other inclusion.

If $S$ is standard then $\Delta(S)$ is standard by (Kechris 1995, Theorem 17.23, 17.24). Since $s \mapsto \delta_s$ is injective, 2 follows from (Kechris 1995, Corollary 15.2).[3]

By 1 we have that for $X \subseteq S$, $\delta(X)$ is $(\Delta(\Sigma)|\delta(S))$-measurable if and only if $X$ is measurable. By using 2, we can state that $\delta(X)$ is $\Delta(\Sigma)$-measurable if and only if $X$ is measurable. □

## 5.2 Non-probabilistic NLMPs

We call a NLMP $\mathbf{S} = (S, \Sigma, \{T_a : a \in L\})$ *non-probabilistic* if for all $a \in L$ and $s \in S$, $T_a(s) \subseteq \delta(S)$. A non-probabilistic NLMP is essentially a labelled transition system (LTS) over a measurable space. However, as we will see, our notions of bisimulation differ from the classical notion for LTS.

---

[3]Alternatively, $(S, \Sigma)$ is isomorphic to $([0,1], \mathbf{B}([0,1]))$ and the functor $\Delta$ can be defined in the category of Polish spaces and continuous functions. It is not hard to show that $\delta$ is a *continuous* embedding. Hence $\delta([0,1])$ is compact in $\Delta([0,1])$ and a fortiori measurable.

We will write $\langle a \rangle Q$ for $\{s : T_a(s) \cap \delta(Q) \neq \varnothing\}$. The interpretation of this is clear: $\langle a \rangle Q$ are the states from which we can reach $Q$ after an $a$-action.

Lemmas 5.2, 5.3, and 5.4 give the formulation of event, state, and traditional bisimulation in the setting of non-probabilistic NLMPs over standard Borel spaces. Lemma 5.1 below, is the basis of the proof of Lemma 5.2 and is also used on the proof of our counterexamples.

**Lemma 5.1.** *Assume $(S, \Sigma)$ is standard and $\Lambda \subseteq \Sigma$ is a sub-$\sigma$-algebra. Let $T_a : S \to \Delta(\Sigma)$ with $T_a(s) \subseteq \delta(S)$ for all $s \in S$. Then $T_a$ is $(\Lambda, H(\Delta(\Lambda)))$-measurable if and only if for all $Q \in \Lambda$, $\langle a \rangle Q \in \Lambda$ (i.e., $\Lambda$ is stable under the mapping $\langle a \rangle \cdot$).*

*Proof.* ($\Rightarrow$) Let $Q \in \Lambda$ and take $\xi = (\Delta(S) \setminus \Delta^{<1}(Q)) \in \Delta(\Lambda)$. Then $\langle a \rangle Q = \{s : T_a(s) \cap \delta(Q) \neq \varnothing\} = \{s : T_a(s) \cap \delta(S) \cap \xi \neq \varnothing\} = T_a^{-1}(H_\xi) \in \Lambda$.

($\Leftarrow$) Let $\xi \in \Delta(\Lambda)$. By Proposition 5.1(1), $\Delta(\Lambda)|\delta(S) = \{\delta(Q) : Q \in \Lambda\}$, and hence $\xi \cap \delta(S) = \delta(Q)$ for some $Q \in \Lambda$. Then $T_a^{-1}(H_\xi) = \{s : T_a(s) \cap \delta(Q) \neq \varnothing\} = \langle a \rangle Q \in \Lambda$. □

Throughout the rest of this section we will assume that $\mathbf{S} = (S, \Sigma, \{T_a : a \in L\})$ is a non-probabilistic NLMP over a standard Borel space. The next lemma is a corollary of Lemma 5.1.

**Lemma 5.2.** *A $\sigma$-algebra $\Lambda \subseteq \Sigma$ is an event bisimulation on $\mathbf{S}$ if and only if it is stable under the mapping $\langle a \rangle \cdot$.*

**Lemma 5.3.** *A symmetric relation $R$ is a state bisimulation on $\mathbf{S}$ if and only if for all $s, t \in S$ such that $s \, R \, t$, it holds that for all $Q \in \Sigma(R)$, $s \in \langle a \rangle Q \Leftrightarrow t \in \langle a \rangle Q$.*

*Proof.* ($\Rightarrow$) Let $s \, R \, t$ and $Q \in \Sigma(R)$. Observe that $\Delta^{\geq 1}(Q) \in \Delta(\Sigma(R))$ and $\Delta^{\geq 1}(Q) \cap \delta(S) = \delta(Q)$. Then $s \in \langle a \rangle Q \Leftrightarrow T(s) \cap \Delta^{\geq 1}(Q) \neq \varnothing \stackrel{\star}{\Leftrightarrow} T(t) \cap \Delta^{\geq 1}(Q) \neq \varnothing \Leftrightarrow t \in \langle a \rangle Q$. In $\star$ we use the fact that $R$ is a state bisimulation.

($\Leftarrow$) Let $s \, R \, t$ and $\xi \in \Delta(\Sigma(R))$. Let $Q$ such that $\delta(Q) = \delta(S) \cap \xi$. Then $T_a(s) \cap \xi = T_a(s) \cap \delta(Q)$ and hence $T_a(s) \cap \xi \neq \varnothing \Leftrightarrow s \in \langle a \rangle Q$. Similarly, $T_a(t) \cap \xi \neq \varnothing \Leftrightarrow t \in \langle a \rangle Q$. If $Q \in \Sigma(R)$, then $s \in \langle a \rangle Q \Leftrightarrow t \in \langle a \rangle Q$ by hypothesis.

We show that indeed $Q \in \Sigma(R)$. Since $\xi \in \Delta(\Sigma(R))$, by Proposition 5.1(2) and (3), $Q \in \Sigma$. It only remains to show that $Q$ is $R$-closed. So, let $x \, R \, y$ and $x \in Q$; hence $\delta_x \in \xi$. But, for any $X \in \Sigma(R)$ and $q \in [0,1]$, $\delta_x \in \Delta^{\geq q}(X) \Leftrightarrow x \in X \Leftrightarrow y \in X \Leftrightarrow \delta_y \in \Delta^{\geq q}(X)$. Since $\delta_x$ and $\delta_y$ cannot be separated by any generator set of $\Delta(\Sigma(R))$, they cannot be separated by a set in $\Delta(\Sigma(R))$ so $\delta_y \in \xi$ and hence $y \in Q$. □

**Lemma 5.4.** *A symmetric relation $R$ is a traditional bisimulation on $\mathbf{S}$ if and only if for all $s, t \in S$ and $\delta_u \in T_a(s)$, if $s \, R \, t$ then there exists $\delta_v \in T_a(t)$ such that $u \, \mathcal{R}(\Sigma(R)) \, v$.*

*Proof.* Assume $s \, R \, t$. Then $T_a(s) \, R \, T_a(t)$ if and only if for every $\mu \in T_a(s)$ there exists $\nu \in T_a(t)$ such that $\mu \, R \, \nu$. But since $\mathbf{S}$ is non-probabilistic, $\mu = \delta_u$, and $\nu = \delta_v$ for some $u, v \in S$. Now $\delta_u \, R \, \delta_v$ means that for every $Q \in \Sigma(R)$, $\delta_u(Q) = \delta_v(Q)$, and this is equivalent to $\forall Q \in \Sigma(R) : u \in Q \Leftrightarrow v \in Q$. The last assertion is $u \, \mathcal{R}(\Sigma(R)) \, v$. □

After the last lemma it should be easy to note that this "measurable" notion of bisimulation is weaker than the standard one for LTS of Milner (1989).

## 5.3 Traditional Bisimilarity $\neq$ Event-, State- Bisimilarity.

Consider the standard Borel space $(S_1, \Sigma_1) = ([0,1] \cup [2,3] \cup \{s,t,x\}, \mathbf{B}([0,1] \cup [2,3] \cup \{s,t,x\}))$ where $\{s,t,x\} \subset \mathbb{R} \setminus [0,3]$. Let $V$ a non-Borel subset of $[2.5, 3]$. Clearly, $[0,1]$ is equinumerous with $[2,3] \setminus V$; pick a bijection $f$ between them. Now, let $L_1 = \{a\} \cup [0,1]$ be the set of labels and let $\mathbf{S_1} = (S_1, \Sigma_1, \{T_a : a \in L_1\})$ where

$$T_a(s) = \delta([2,3])$$
$$T_a(t) = \delta([0,1])$$
$$T_r(r) = T_r(f(r)) = \{\delta_x\} \qquad \text{if } r \in [0,1]$$
$$T_c(y) = \varnothing \qquad \text{otherwise.}$$

Now, take $\mathcal{F}$ to be $\{\{s,t\}, \{x\}, \{r, f(r)\}_{r \in [0,1]}\}$ and $R = \mathcal{R}(\sigma(\mathcal{F}))$.

**Lemma 5.5.** $\mathbf{S_1}$ *is a non-probabilistic NLMP, $\sigma(\mathcal{F})$ is an event bisimulation and $R$ is a state bisimulation.*

*Proof.* First, notice that for all $c, y$, $T_c(y) \in \Delta(\Sigma_1)$ by Proposition 5.1(3). The proof that $T_c$ is a measurable map for each $c \in L_1$ is routine.

We check that $\sigma(\mathcal{F})$ is an event bisimulation. Observe first that for all $Q \in \sigma(\mathcal{F})$, $\langle a \rangle Q$ is empty or equal to $\{s,t\} \in \sigma(\mathcal{F})$, and hence $\sigma(\mathcal{F})$ is stable under $T_a$ by Lemma 5.1. For $0 \leq r \leq 1$, $\langle r \rangle Q \neq \varnothing$ if and only if $x \in Q$, and in that case

$$\langle r \rangle Q = \{r, f(r)\} \in \sigma(\mathcal{F}). \tag{1}$$

Now, we show that $R$ is a state bisimulation. By way of contradiction, using Lemma 5.3, assume that there exists $Q \in \Sigma_1(R)$, $c \in L_1$ and $z, y \in S_1$ such that $z \, R \, y$ and $z \in \langle c \rangle Q$ but $y \notin \langle c \rangle Q$. Hence $\langle c \rangle Q$ must not be $R$-closed. By the preceding calculation (1), for $0 \leq r \leq 1$ and *every* $Q \in \Sigma_1$, $\langle r \rangle Q$ is $R$-closed. Then, it should be the case that $\langle a \rangle Q$ is not $R$-closed. Observe that the only $R$-closed sets $Q \subseteq S_1$ such that $\langle a \rangle Q$ is not $R$-closed are of the form $A \cup V$ where $A \in \{\varnothing, \{s,t,x\}, \{s,t\}, \{x\}\}$. This set $Q$ is non-measurable since $V$ was chosen to be not measurable. But then $Q$ is not in $\Sigma_1(R)$, an absurdity. $\square$

**Theorem 5.6.** *State bisimilarity (resp. event bisimilarity) and traditional bisimilarity differ in $\mathbf{S_1}$.*

*Proof.* Because of Lemma 5.5, it suffices to show that $s$ and $t$ are not traditionally bisimilar.

It is easy to see that for $0 \leq r \leq 1$, $r \not\sim_t y$ if $y \notin \{r, f(r)\}$: we have $T_r(r)$ nonempty but $T_r(y) = \varnothing$. Therefore $\{r, f(r)\}$ is $\sim_t$-closed for every $0 \leq r \leq 1$ and hence $\{r, f(r)\} \in \Sigma_1(\sim_t)$.

By way of contradiction, now assume $s \sim_t t$. Let $y \in V$. Since $\delta_y \in T_a(s)$, by Lemma 5.4, there must exist some $0 \leq r \leq 1$ such that $y \, \mathcal{R}(\Sigma_1(\sim_t)) \, r$. But $y > 1$ and is not in the image of $f$, hence the set $\{r, f(r)\} \in \Sigma_1(\sim_t)$ separates $y$ from $r$. This contradicts the fact that $y \, \mathcal{R}(\Sigma_1(\sim_t)) \, r$.

Since $\sim_s \subseteq \sim_e$, event bisimilarity and traditional bisimilarity also differ in $\mathbf{S_1}$. $\square$

## 5.4 State Bisimilarity $\neq$ Event Bisimilarity

In this last section, we prove that the greatest event bisimulation $\sim_e$ is not contained in $\sim_s$. We do this by slightly modifying $\mathbf{S_1}$. We now take $V$ to be the interval $(2.5, 3]$ and let $(S_2, \Sigma_2) = (S_1, \Sigma_1)$. We complete the construction of a non-probabilistic NLMP by picking any bijection $f$ between $[0, 1]$ and $[2, 2.5]$. The transition is defined just like for $\mathbf{S_1}$ only that using the the new $f$. We also use family $\mathcal{F}$ but defined with the new $f$.

**Lemma 5.7.** $V \notin \sigma(\mathcal{F})$.

*Proof.* It is clear that every member of $\sigma(\mathcal{F})$ is countable or has a countable complement, from which the lemma follows. □

The proof of Lemma 5.5 works equally fine for the following lemma.

**Lemma 5.8.** $\mathbf{S_2} = (S_2, \Sigma_2, \{T_a : a \in L_1\})$ is a non-probabilistic NLMP and $\sigma(\mathcal{F})$ is an event bisimulation.

In this case, relation $R = \mathcal{R}(\sigma(\mathcal{F}))$ is an event bisimulation that it is not a state bisimulation.

**Theorem 5.9.** *Event and state bisimilarity differ in $\mathbf{S_2}$.*

*Proof.* Since $(s, t) \in R \subseteq \sim_e$, we just have to show that $s \not\sim_s t$. Observe that $V \in \Sigma_2(R)$. If $s$ and $t$ were state-bisimilar, by Lemma 5.3, it would be the case that $s \in \langle a \rangle V$ iff $t \in \langle a \rangle V$. But this is absurd since $\delta_3 \in T_a(s) \cap \delta(V)$ and $T_a(t) \cap \delta(V) = \varnothing$. □

## 6 Concluding remarks

In order to define a process theory that permits the verification of compositionally modeled systems against simple (may be nondeterministic) specifications, it is necessary to have at hand a semantic relation that allows for abstraction such as weak bisimulation. In this setting, internal nondeterminism is crucial.

In this paper we introduced the model of nondeterministic labeled Markov processes that allows for the modeling of continuous probabilistic systems with internal nondeterminism. Contrarily to similar models (D'Argenio 1999; Bravetti 2002; Bravetti and D'Argenio 2004; D'Argenio and Katoen 2005; Cattani 2005), NLMPs are defined to have a measure theoretic structure. In particular, we require that the transition relation is a measurable function that maps on measurable sets. This was devised so that it is possible to build the rest of the theory (particularly event bisimulation and logic, but also schedulers are definable). We have shown that NLMPs extend naturally LMPs. For the definition of the transition and the development of the whole work, Def. 3.2 is crucial, as it provides the foundation for dealing with nondeterminism.

As a first step towards the desired process theory, we gave different definitions of bisimulations. We proposed three possible generalizations of the two bisimulations on LMPs. The event bisimulation responds exactly to the same definition principle both in LMP and NLMP. Instead, the state bisimulation in LMPs generalizes to NLMPs as state bisimulation and as traditional bisimulation. We know that traditional bisimulation is finer than state bisimulation and, in Theorems 3.6 and 3.7, we gave sufficient conditions under which they agree.

We also gave a logical characterization of event bisimulation (Theorem 4.5). Such logic ($\mathcal{L}$) can be seen as a revision of the one introduced by Parma and Segala (2007) in a discrete probabilistic setting. Formulas in our setting belong to two different classes: state formulas and measure formulas. Notice that negation and infinitary (but denumerable) disjunction (or conjunction) is only present on the second class, meaning that the complexity of the model lies precisely on the internal nondeterminism.

A consequence of the characterization is that the logic is sound for state and traditional bisimulations (Theorem 4.6). For the restricted case of image finite NLMPs running on analytic Borel spaces, all equivalences coincide (Theorem 4.10). Notice that the logic we used to show such equivalence is in fact a sublogic of $\mathcal{L}$ which has already appeared in a preliminary work (Celayes 2006).

The coincidence among all equivalence does not generalise to arbitrary NLMPs as we have shown in Theorems 5.6 and 5.9. Observe that the counterxamples presented in these theorems are non-probabilistic NLMPs over standard Borel spaces with uncountable branching. Somehow, this shows that Theorems 3.6 and 3.7 are the best possible to equate traditional and state bisimulation, even if we assume that the state space is the Borel space of the real numbers. Though we did not present a theorem, we mentioned a third important difference on these "measure-theoretic" LTSs: in the general case, Park-Milner's bisimulation is strictly finer than traditional bisimulation. The last one considers the measure space of the state space, while the first one does not (or, alternatively, it only considers the discrete $\sigma$-algebra $2^S$).

Some additional observations on the counterexamples are in order. First, counterexample $\mathbf{S_1}$ in Theorem 5.6 relies on the fact that state bisimulation cannot distinguish a non-measurable set $V$ while traditional bisimulation can. In our point of view, such distinction should not be possible since $V$ has no measure. Second, counterexample $\mathbf{S_2}$ in Theorem 5.9 makes a difference on measurable set $V$ that the event bisimulation cannot distinguish. In our opinion, such distinction should be observed since a possible scheduler may lead to such set of states with certain probability. Notice that in this example, states in $V$ do not allow the system to reach state $x$ from $s$, while $x$ can always be reached from $t$. In this sense, we argue that state bisimulation is the most appropriate definition.

Somehow, this is dissapointing since logic $\mathcal{L}$ has a natural definition but, as it completely characterizes event bisimulation, it will not be able to test the presence of states like those in $V$ in $\mathbf{S_2}$. This is due to the fact that the logic cannot test transitions bearing continuously many labels. This calls for adding structure to the set of labels on the NLMP. In any case, this would also be necessary for the definition of schedulers and probabilistic trace semantics.

At the moment, we are indeed busy on the definition of NLMPs with labels equipped with a $\sigma$-algebra, as well as on the study of schedulers for these objects and probabilistic trace semantics. This will allow us to contrast the two local behavioral equivalences, state and traditional bisimulation. It is expected that at least one of them implies a global behavioral equivalence, like probabilistic trace equality. Schedulers would also let us define probabilistic weak transitions and their related bisimulations. We are also busy on trying to refine the idea of event bisimulation and the logic so that they can distinguish situations like the one shown by NLMP $\mathbf{S_2}$.

If necessary, we will restrict only to standard Borel spaces. Confining to standard Borel spaces is not as restricting as it seems since most natural problems arise in this setting. For example, we have shown elsewhere that the underlying semantics of stochastic automata (D'Argenio 1999) in terms of NLMP meets most of the restrictions required in this article: it runs on standard Borel spaces and it is image finite. We recall that stochastic automata and similar models are used to give semantics to stochastic process algebras and specification languages (D'Argenio 1999; Bravetti 2002; Bravetti and D'Argenio 2004; D'Argenio and Katoen 2005; Bohnenkamp et al. 2006, etc.) which, in turn, are used to model dynamic systems. Moreover, LMP-like models restricted to standard Borel spaces have been studied by Doberkat (2007).

*Acknowledgments:* The logic presented here has its roots on Celayes's (2006) master thesis which has been jointly supervised by the first two coauthors of this article. We thank Pablo Celayes for his initial colaboration. We also thank Ignacio Viglizzo for fruitful discussions. In particular he pointed out the connection of Def. 3.2 to hit-and-miss topologies.